\documentclass[12pt]{article}
\usepackage{enumerate}
\usepackage{amsfonts}
\usepackage{amsmath}
\usepackage{amssymb}
\usepackage{amsthm}

\def\H{\mathcal{H}}

\def\S{\mathfrak{S}}

\def\T{\mathfrak{T}}
\def\B{\mathfrak{B}}

\newcommand{\Tr}{\mathrm{Tr}}

\newcommand{\ri}{\mathrm{ri}}

\newcounter{defin}  \newcounter{lemma}  \newcounter{theorem}
\newcounter{property} \newcounter{corol}  \newcounter{remark} \newcounter{example}

\newenvironment{lemma}{\par\refstepcounter{lemma}
     \textbf{Lemma \thelemma.} }{\rm\par}

\newenvironment{property}{\par\refstepcounter{property}
     \textbf{Proposition \theproperty.}\ }{\rm\par}
\newenvironment{corollary}{\par\refstepcounter{corol}
     \textbf{Corollary \thecorol.} }{\rm\par}

\newenvironment{remark}{\par\refstepcounter{remark}
     \textbf{Remark \theremark.}}{\rm\par}

\begin{document}
\title{On singular Bosonic linear channels\thanks{This work was
partially supported by the
program ``Mathematical control theory and dynamical systems'' of RAS and by the RFBR grants 12-01-00319a and 13-01-00295a.}}
\author{M.E. Shirokov\footnote{email:msh@mi.ras.ru}\\
Steklov Mathematical Institute, RAS, Moscow}
\date{}
\maketitle 
\begin{abstract}
Properties of Bosonic linear (quasi-free) channels, in particular, Bosonic Gaussian channels with two types of
degeneracy are considered.

The first type of degeneracy can be interpreted as existence of
noise-free canonical variables (for Gaussian channels it means that $\det\alpha=0$).
It is shown that this
degeneracy implies existence of (infinitely many) "direct
sum decompositions" of  Bosonic linear channel, which clarifies
reversibility properties of this channel (described in arXiv:1212.2354) and
provides explicit construction of reversing channels.

The second type of degeneracy consists in rank deficiency of the
operator describing transformations of canonical variables. It is
shown that this degeneracy implies  existence of (infinitely many)
decompositions of input space into direct sum of orthogonal
subspaces such that the restriction of  Bosonic linear channel to
each of these subspaces is a discrete classical-quantum channel.
\end{abstract}
\vspace{15pt}

\section{Preliminaries} Let $\H$ be a separable Hilbert space, $\B(\H)$ and
$\mathfrak{T}( \mathcal{H})$ -- the Banach spaces of all bounded
operators in $\mathcal{H}$ and of all trace-class operators in $\H$
correspondingly, $\S(\H)$ -- the closed convex subset of
$\mathfrak{T}( \H)$ consisting of positive operators with unit trace
called \emph{states} \cite{H-SCI}.

Denote by $I_{\mathcal{H}}$ and $\mathrm{Id}_{\mathcal{H}}$ the
unit operator in a Hilbert space $\mathcal{H}$ and the identity
transformation of the Banach space $\mathfrak{T}(\mathcal{H})$
correspondingly.\smallskip

A  completely positive trace preserving linear map
$\Phi:\mathfrak{T}(\mathcal{H}_A)\rightarrow\mathfrak{T}(\mathcal{H}_B)$
is called  \emph{quantum channel} \cite{H-SCI}.
\medskip

A channel $\Phi:\T(\H_A)\rightarrow\T(\H_B)$ is called
\emph{classical-quantum of discrete type} (briefly, \emph{discrete
c-q channel}) if it has the following representation
\begin{equation}\label{c-q-rep}
\Phi(\rho)=\sum_{k=1}^{\dim\H_A}\langle k|\rho|k\rangle\sigma_k,\quad \rho\in \T(\H_A),
\end{equation}
where $\{|k\rangle\}$ is an orthonormal basis in $\H_A$ and
$\{\sigma_k\}$ is a collection of states in $\S(\H_B)$.\footnote{We
use the term "discrete" here, since in infinite dimensions there
exist channels naturally called classical-quantum, which has no
representation   (\ref{c-q-rep}) \cite{H-CQC}.}
\medskip

Let $\mathcal{H}_{X}$ $(X=A,B,...)$ be the space of irreducible representation of
the Canonical Commutation Relations (CCR)
\begin{equation*}
W_X(z)W_X(z^{\prime })=\exp \left[-\frac{\mathrm{i}}{2}\,z^{\top }\Delta
_{X}z^{\prime }\right] W_X(z^{\prime }+z)  
\end{equation*}
with a symplectic space $(Z_{X},\Delta _{X})$ and the Weyl operators
$W_{X}(z)$ \cite[Ch.12]{H-SCI}. Denote by $s_X$ the number of modes
of the system $X$, i.e. $2s_X=\dim Z_X$.\smallskip

We will use the Schrodinger representation of CCR: for a given
symplectic basis $\{e_i, h_i\}$ in $Z_X$, we can identify the space
$\H_X$ with the space $L_2(\mathbb{R}^{s_X})$ of complex-valued
functions of $s_X$ variables (which will be denoted
$\xi_1,...,\xi_{s_X}$) and the Weyl operators $W_X(e_i)$ and
$W_X(h_i)$ with the operators
$$
\psi(\xi_1,...,\xi_{s_X})\mapsto e^{\mathrm{i}\xi_i}
\psi(\xi_1,...,\xi_{s_X})\;\;\,\text{and}\;\;
\psi(\xi_1,...,\xi_{s_A})\mapsto
\psi(\xi_1,...,\xi_{i}+1,...,\xi_{s_X}).
$$

A Bosonic linear channel  $\Phi
:\mathfrak{T}(\mathcal{H}_{A})\rightarrow
\mathfrak{T}(\mathcal{H}_{B})$ is defined via the action of its dual
$\Phi^{\ast }:\mathfrak{B}(\mathcal{H}_{B})\rightarrow
\mathfrak{B}(\mathcal{H}_{A})$ on the Weyl operators:
\begin{equation}
\Phi^{\ast}(W_{B}(z))=W_A(Kz)f(z),\quad z\in Z_B, \label{blc}
\end{equation}
where $K$ is a linear operator $Z_{B}\rightarrow Z_{A}$, and $f(z)$
is a complex continuous function on $Z_B$ such that $\,f(0)=1\,$ and
the matrix with the elements\break
$f(z_s-z_r)\exp\left(\,\frac{\mathrm{i}}{2}\,z_s^{\top }[\Delta_B-K^{\top }\Delta_A K]
z_r\right)$ is positive for any finite subset $\{z_s\}$ of $Z_B$
\cite{DVV,H-BC,H-SCI}.\footnote{In \cite{DVV} this channel is called quasi-free.}
This channel will be denoted $\Phi_{K,f}$.\smallskip

A very important class of Bosonic linear channels consists of
Bosonic Gaussian channels defined by (\ref{blc}) with the Gaussian
function
$$
f(z)=\exp \left[\,\mathrm{i}l^{\top }z-\textstyle\frac{1}{2} z^{\top }\alpha
z\,\right],
$$
where $l\,$ is a $\,2s_B$-dimensional real row and $\,\alpha\,$ is a real
symmetric $\,(2s_B)\times(2s_B)$ matrix satisfying the inequality
\begin{equation*}
\alpha \geq \pm \frac{\mathrm{i}}{2}\left[\, \Delta _{B}-K^{\top}\Delta_{A}K\,\right].
\end{equation*}
Bosonic Gaussian channels play a central role in
infinite-dimensional quantum information theory \cite{E&W,H-SCI}.\smallskip

Denote by $Z_f$ the subset $f^{-1}(1)=\{z\in
Z_B\,|\,f(z)=1\}$. One
can show that $\,Z_f$ is a linear subspace of $\,Z_B$ coinciding with
$\,\ker\alpha\,$ in the case of Gaussian function $f$
\cite[Ch.12]{H-SCI}.\smallskip

In this paper we consider properties of a Bosonic
linear channel $\Phi_{K,f}$ with the following two types of degeneracy:
\begin{itemize}
\item $Z_f\doteq f^{-1}(1)\neq\{0\}$;
\item $\mathrm{rank}K<\dim Z_A\;$ ($\mathrm{Ran}K\neq Z_A$).
\end{itemize}
These types of degeneracy are related via the notion of a weak
complementary channel (see detailed definition in
\cite[Ch.6]{H-SCI}). Indeed, under the assumption of existence of
Bosonic linear unitary dilation\footnote{A sufficient condition for
existence of such dilation is given in \cite{DVV}. For Bosonic
Gaussian channels it is proved in \cite{Caruso} (see also
\cite[Ch.12]{H-SCI}).} for a channel $\Phi_{K,f}$ a weak
complementary channel to $\Phi_{K,f}$ is a Bosonic linear channel
$\Phi_{L,g}$ from $\T(\H_A)$ into $\T(\H_E)$, where $E$ is a Bosonic
system-environment \cite{Caruso,H-SCI,BRC}. Lemma 2 in \cite{BRC}
implies\footnote{Here $"\perp"$ denotes the skew-orthogonal
complementary subspace. We will always use this sense of the symbol
$"\perp"$ dealing with a subspace of a symplectic space.}
$$
\dim[\mathrm{Ran}L]^{\perp}=\dim Z_f,\quad \dim[\mathrm{Ran}K]^{\perp}=\dim Z_g.
$$
Hence a channel $\Phi_{K,f}$ has the first type of degeneracy if and
only if any\footnote{In contrast to complementary channel a weak
complementary channel is not uniquely defined.} weak complementary
channel to $\Phi_{K,f}$ has the second type of degeneracy and vice
versa.

\section{The case $Z_f\doteq f^{-1}(1)\neq\{0\}\,$ ($\det\alpha=0$)}

Physically, the condition $Z_f\neq\{0\}$ means (in the Heisenberg
picture)  that the channel $\Phi_{K,f}^{*}$ injects no noise
in some canonical variables of the system $B$ (which can be called
noise-free canonical variables).\smallskip

We will use the following simple observation (see e.g. Lemma 2 in
\cite{BRC}).\smallskip
\begin{lemma}\label{main-l}
\emph{The restriction of the operator $K$ to the subspace $Z_f$ is
non-degenerate and $\,\Delta_A(Kz_1,Kz_2)=\Delta_B(z_1,z_2)$ for all
$\,z_1,z_2\in Z_f$.}
\end{lemma}\smallskip

Let $Z_{A_0}$ and $Z_{B_0}$ be minimal symplectic subspaces
containing respectively $K(Z_f)$ and $Z_f$. By Lemma \ref{main-l}
$\dim Z_{A_0}=\dim Z_{B_0}$. We have
$$
Z_X=Z_{X_0}\oplus Z_{X_*}\;\,(Z_{X_*}=[Z_{X_0}]^{\perp}),\quad
\H_X=\H_{X_0}\otimes\H_{X_*},\quad (X=A,B).
$$

Since $W_B(z)=W_{B_0}(z)\otimes I_{B_*}$ and
$W_A(Kz)=W_{A_0}(Kz)\otimes I_{A_*}$ for all $z\in Z_f$, the von
Neumann algebras $\mathcal{A}$ and $\mathcal{B}$ generated
respectively by the families  $\{W_A(Kz)\}_{z\in Z_f}$ and
$\{W_B(z)\}_{z\in Z_f}$ have the following forms
$$
\mathcal{A}=\mathcal{A}_0\otimes I_{A_*},\quad
\mathcal{B}=\mathcal{B}_0\otimes I_{B_*},
$$
where $\mathcal{A}_0$ and $\mathcal{B}_0$ are algebras acting
respectively on $\H_{A_0}$ and on $\H_{B_0}$.\smallskip

By Lemma \ref{main-l} and Lemma \ref{last-l} in the Appendix there
exists a symplectic transformation  $T:Z_{B_0}\rightarrow Z_{A_0}$
such that $Kz=Tz$ for all $z\in Z_f$. Hence $\,W_{A_0}(Kz)=U_T
W_{B_0}(z)U_T^*\,$ for all $z\in Z_f$, where $U_T$ is the unitary operator implementing
$T$. It follows that  the algebras $\mathcal{A}_0$ and
$\mathcal{B}_0$ are unitary equivalent, i.e. $\mathcal{A}_0=U_T
\mathcal{B}_0 U_T^*$.\smallskip

Since $\Phi_{K,f}^*(W_{B_0}(z)\otimes I_{B_*})=W_{A_0}(Kz)\otimes
I_{A_*}$ for  all $z\in Z_f$ (by definition of the channel
$\Phi_{K,f}$), the restriction of the dual channel $\Phi^*_{K,f}$ to
the algebra $\mathcal{B}$ coincides with the isomorphism
$$
\mathcal{B}=\mathcal{B}_0\otimes I_{B_*}\ni Y\otimes I_{B_*}\,\mapsto\,
\left[U_TYU_T^*\right]\otimes I_{A_*}\in\mathcal{A}_0\otimes
I_{A_*}=\mathcal{A}
$$
between the algebras $\mathcal{B}$ and $\mathcal{A}$.\smallskip

It follows, in particular, that for an arbitrary projector
$P\in\mathcal{A}$ there exists a projector $Q\in\mathcal{B}$ such
that $P=\Phi^*_{K,f}(Q)$. It is easy to see that this relation means
that $\Phi_{K,f}\left(\T(P(\H_A))\right)\subseteq\T(Q(\H_B))$. So,
we obtain the following observation.\pagebreak

\begin{property}\label{d-s-dec}
\emph{Let $\,\Phi_{K,f}$ be a  Bosonic  channel with
$\,Z_f\neq\{0\}$. For an arbitrary orthogonal resolution of the
identity \footnote{An orthogonal resolution of the identity is a
family of mutually orthogonal projectors whose sum coincides with
the identity operator.}
$\{P_k\}\subset\mathcal{A}\doteq\left[\{W_A(Kz)\}_{z\in
Z_f}\right]''$ there exists an orthogonal resolution of the identity
$\{Q_k\}\subset\mathcal{B}\doteq\left[\{W_B(z)\}_{z\in
Z_f}\right]''$ such that $P_k=\Phi^*_{K,f}(Q_k)$ for all $\,k$ and
hence
\begin{equation}\label{i-r}
\Phi_{K,f}\left(\T(\H_A^k)\right)\subseteq\T(\H_B^k)\quad\forall k,
\end{equation}
where $\,\H_A^k=P_k(\H_A)$ and $\,\H_B^k=Q_k(\H_B)$ (so that
$\,\H_X=\bigoplus_k\H_X^k,X=A,B$).}
\end{property}
\medskip

\begin{remark}\label{d-s-dec-r}
If $Z^c_f$ is a nontrivial isotropic subspace of $Z_f$ then the
algebra $\mathcal{A}^c=[\{W_A(Kz)\}_{z\in Z^c_f}]''$ is commutative
and isomorphic to the algebra $L_{\infty}(\mathbb{R}^d)$, where
$d=\dim Z^c_f$. It follows that any element of an orthogonal
resolution of the identity
$\{P_k\}\subset\mathcal{A}^c\subseteq\mathcal{A}$ can be represented
as a sum of mutually orthogonal  projectors in $\mathcal{A}^c$.
Hence, by Proposition \ref{d-s-dec}, any subspace of the
corresponding decomposition $\H_A=\bigoplus_k\H_A^k$ can be, in
turn, decomposed into direct sum of orthogonal subspaces, for each
of which the invariance relation similar to (\ref{i-r}) holds.
\end{remark}
\medskip
A quantum channel
$\Phi:\mathfrak{T}(\mathcal{H}_A)\rightarrow\mathfrak{T}(\mathcal{H}_B)$
is called \emph{reversible} (or sufficient) with respect to a family $\S$ of
states in $\mathfrak{S}(\mathcal{H}_A)$ if there exists a quantum
channel
$\Psi:\mathfrak{T}(\mathcal{H}_B)\rightarrow\mathfrak{T}(\mathcal{H}_A)$
such that $\Psi(\Phi(\rho))=\rho$ for all $\rho\in\S$
\cite{J&P,J-rev}.\smallskip

The above family $\S$ is naturally called \emph{reversed family} for the
channel $\Phi$, while the channel $\Psi$ may be called \emph{reversing channel}.
\smallskip

Necessary and sufficient conditions for reversibility of Bosonic
linear channels with respect to orthogonal and non-orthogonal families
of pure states (as well as explicit forms of reversed families)
are explored in \cite{BRC} by using the "method of complementary
channel".

Proposition \ref{d-s-dec} clarifies the sufficiency in the
"orthogonal part" of these conditions. Moreover, it shows
reversibility of the channel $\Phi_{K,f}$ such that $Z_f\neq\{0\}$ with
respect to particular orthogonal families of states (not necessarily
pure) and provides explicit description of reversing channels.
\smallskip

\begin{corollary}\label{d-s-dec-c}
\emph{Let $\,\Phi_{K,f}$ be a  Bosonic linear channel such that
$Z_f\neq\{0\}$ and $\{P_k\}$ an orthogonal resolution of the identity in $\mathcal{A}\doteq\left[\{W_A(Kz)\}_{z\in Z_f}\right]''$.
The channel $\,\Phi_{K,f}$ is reversible with respect to any
family $\{\rho_k\}$ of states in $\,\S(\H_A)$ such that
$\,\mathrm{supp}\rho_k\subseteq \H_A^k=P_k(\H_A)$. The simplest
reversing channel for the family $\{\rho_k\}$ has the form}
$$
\Psi(\sigma)=\sum_k[\Tr Q_k\sigma]\rho_k,\quad  \sigma\in\S(\H_B),
$$
\emph{where $\{Q_k\}$ is the orthogonal resolution of the identity
in $\,\mathcal{B}\doteq\left[\{W_B(z)\}_{z\in Z_f}\right]''$
described in Proposition \ref{d-s-dec}.}
\end{corollary}\smallskip

Corollary \ref{d-s-dec-c} gives an explicit proof of the part of
Theorem 2 in \cite{BRC}, which states reversibility of the channel
$\Phi_{K,f}$ with respect to non-complete orthogonal families of
pure states provided that $Z_f$ is a nontrivial isotropic subspace
of $Z_B$. It also shows sufficiency of the condition obtained in
Section 4.3 in \cite{BRC} describing all reversed families in this
case.\smallskip

Indeed, if $Z_f$ is a nontrivial isotropic subspace of $Z_B$ then
the above-defined algebras $\mathcal{A}=\mathcal{A}_0\otimes
I_{A_*}$ and $\mathcal{B}=\mathcal{B}_0\otimes I_{B_*}$ are
commutative and in the Schrodinger representation (described in
Section 1) $\mathcal{A}_0\cong\mathcal{B}_0\cong
L_{\infty}(\mathbb{R}^d)$, where $d=\dim Z_f$. For $X=A,B\,$ the
algebra $\mathcal{X}=\mathcal{X}_0\otimes I_{X_*}$ acts on the space
$\H_X\cong L_{2}(\mathbb{R}^{s_X})$ as follows:
\begin{equation}\label{f}
\!(F\otimes
I_{X_*}\psi)(\xi_1,\ldots,\xi_{s_X})=F(\xi_1,\ldots,\xi_{d})\psi(\xi_1,\ldots,\xi_{s_X}),\,
F\in \mathcal{X}_0\cong L_{\infty}(\mathbb{R}^d).\!\!
\end{equation}
Since projectors in $L_{\infty}(\mathbb{R}^d)$ correspond to
indicator functions of subsets of $\mathbb{R}^d$, any orthogonal
resolutions of the identity $\{P_k\}$ and $\{Q_k\}$ involved in Proposition
\ref{d-s-dec} correspond to a decomposition $\{D_k\}$ of
$\mathbb{R}^d$ into disjoint measurable subsets. So,
$\H_A^k=P_k(\H_A)=L_{2}(D_k\times\mathbb{R}^{s_A-d})$ is the
subspace of $\H_A=L_{2}(\mathbb{R}^{s_A})$ consisting of functions
vanishing almost everywhere outside the cylinder
$$
D_k\times\mathbb{R}^{s_A-d}=\{(\xi_1,\ldots,\xi_{s_A})\,|\,(\xi_1,\ldots,\xi_{d})\in
D_k\}.
$$
while $\H_B^k=Q_k(\H_B)=L_{2}(D_k\times\mathbb{R}^{s_B-d})$ is the
subspace of $\H_B=L_{2}(\mathbb{R}^{s_B})$ consisting of functions
vanishing almost everywhere outside the cylinder
$$
D_k\times\mathbb{R}^{s_B-d}=\{(\xi_1,\ldots,\xi_{s_B})\,|\,(\xi_1,\ldots,\xi_{d})\in
D_k\}.
$$
Proposition \ref{d-s-dec} asserts that all states supported by
$\H_A^k$ are transformed by the channel $\Phi_{K,f}$ into states
supported by $\H_B^k$.\smallskip

It follows (as stated in  Corollary \ref{d-s-dec-c}) that any family
$\{|\psi_k\rangle\langle\psi_k |\}$ such that $\psi_k \in
L_{2}(D_k\times\mathbb{R}^{s_A-d})$ for each $k$ is a reversed family for the
channel $\Phi_{K,f}$ and the role of reversing channel is played by the
map $\sigma\mapsto\sum_k[\Tr
Q_k\sigma]|\psi_k\rangle\langle\psi_k |$. The arguments from
Subsection 4.3 in \cite{BRC} (the case $\ri(\Phi_{K,f})=01$) show
that \emph{all} reversed families for the
channel $\Phi_{K,f}$ have the above-described form.

\section{The case $\mathrm{rank}K<\dim Z_A$}

It is shown in \cite[Proposition 3]{BRC} that the condition
$\mathrm{rank}K<\dim Z_A$ is equivalent to existence of discrete c-q
subchannels of the channel $\Phi_{K,f}$. The following proposition
strengthens this observation.\smallskip

\begin{property}\label{c-q-dec} \emph{If $\,\mathrm{rank}K<\dim Z_A$
then for arbitrary given $m\leq+\infty$ there exists a decomposition
$\H_A=\bigoplus_{k=1}^{+\infty}\H_A^k$ such that $\,\dim\H_A^k=m$  and
$$
\Phi_{K,f}|_{\T(\H_A^k)}\textit{\;\,is a discrete c-q channel for
each}\;\,k.
$$
Any such decomposition is determined by the following parameters:
\begin{itemize}
    \item non-trivial isotropic subspace
$Z_0\subseteq[\mathrm{Ran}K]^{\perp}$;
    \item non-degenerate
decomposition $\{D_i\}_{i=1}^m$ of $\,\mathbb{R}^d$, where $d=\dim
Z_0$;
    \item collection $\{E_i\}_{i=1}^m$, where $E_i=\{|e^i_{k}\rangle\}_{k=1}^{+\infty}$ is an ONB
    in $L_{2}(D_i\times\mathbb{R}^{s_A-d})$;
    \item collection $\{\pi_i\}_{i=1}^m$, where $\pi_i$ is a permutation of
    $\,\mathbb{N}$.
\end{itemize}
For a given choice of these parameters,
\begin{equation}\label{s-c-q-rep}
\Phi_{K,f}(\rho)=\sum_{i=1}^m\langle
e^i_{\pi_i(k)}|\rho|e^i_{\pi_i(k)}\rangle\sigma^i_k,\quad
\rho\in\S(\H_A^k),
\end{equation}
for each $k$, where $\{\sigma^i_k\}$ is a collection of states in
$\H_B$.}\medskip
\end{property}

\textbf{Proof.} Let $Z_0$ be a non-trivial isotropic subspace of
$[\mathrm{Ran}K]^{\perp}$. Then the commutative von Neumann algebra
$\mathcal{A}_0=\left[\{W_A(z)\}_{z\in Z_0}\right]''$ is contained in
the algebra $\left[\{W_A(Kz)\}_{z\in Z_B}\right]'$. In the
Schrodinger representation the algebra $\mathcal{A}_0$ coincides
with the algebra $L_{\infty}(\mathbb{R}^d)$, where $d=\dim Z_0$,
acting on the space $\H_A=L_{2}(\mathbb{R}^{s_A})$ in accordance
with formula (\ref{f}). Hence an arbitrary orthogonal resolution of
the identity $\{P_i\}_{i=1}^m$ in $\mathcal{A}_0$ corresponds to a
decomposition $\{D_i\}_{i=1}^m$ of $\,\mathbb{R}^d$ into disjoint
measurable subsets (in the sense that the projector $P_i$
corresponds to the indicator function of the set $D_i$).

For a given decomposition $\{D_i\}_{i=1}^m$ of $\,\mathbb{R}^d$ choose collections $\{E_i\}_{i=1}^m$ and $\{\pi_i\}_{i=1}^m$, where
$E_i=\{|e^i_{k}\rangle\}_{k=1}^{+\infty}$ is an orthonormal basis in
$L_{2}(D_i\times\mathbb{R}^{s_A-d})=P_i(\H_A)$ and $\pi_i$ is a
permutation of $\mathbb{N}$. For each $k\in\mathbb{N}$ let $\H_A^k$
be the subspace of $\H_A$ generated by the family
$\{|e^i_{\pi_i(k)}\rangle\}_{i=1}^{m}$. Since
$\{P_i\}_{i=1}^m\subset\left[\{W_A(Kz)\}_{z\in Z_B}\right]'$, we have
$$
\langle e^i_{\pi_i(k)}|W_A(Kz)|e^j_{\pi_j(k)}\rangle=0\quad \textup{for all}\;\,
i\neq j \;\,\textup{and  all}\;k.
$$
By Lemma 3 in \cite[Appendix 6.1]{BRC} the subchannel of the
channel $\Phi_{K,f}$ corresponding to the subspace $\H_A^k$ is a
discrete c-q channel for each $k$ having representation
(\ref{s-c-q-rep}). $\square$ \medskip

\begin{corollary}\label{c-q-dec-c}
\emph{Let $\H_A=\bigoplus_{k=1}^{+\infty}\H_A^k$ be a given
decomposition from Proposition \ref{c-q-dec} and $P_k$  the
projector onto $\H_A^k$ for each $k$. The channel $\,\Phi_{K,f}$
coincides with the discrete c-q channel
\begin{equation*}
\rho\,\mapsto\,\sum_{k=1}^{+\infty}\sum_{i=1}^m\langle
e^i_{\pi_i(k)}|\rho|e^i_{\pi_i(k)}\rangle\sigma^i_k
\end{equation*}
on the set $\,\{\rho\in\S(\H_A)\,|\,\rho=\Pi(\rho)\}$, where $\Pi(\rho)=\sum_{k=1}^{+\infty}P_k\rho P_k$.}
\end{corollary}

\medskip
Corollary \ref{c-q-dec-c} implies, in particular, that
$\Phi_{K,f}\circ\Pi$ is a discrete c-q channel.

\section*{Appendix}

\begin{lemma}\label{last-l}
\emph{Let $\,Z_0$ be an arbitrary subspace of $\,Z$ and
$\,K:Z_0\rightarrow Z$ a linear map such that $\,Kz_1\neq Kz_2$ and
$\,\Delta(Kz_1,Kz_2)=\Delta(z_1,z_2)$ for all $\,z_1\neq z_2$ in $Z_0$.
There exists  a symplectic transformation $\,T:Z\rightarrow Z$ such
that $\,Kz=Tz$ for all $z\in Z_0$.}\footnote{I would be grateful for a direct reference on this result.}
\end{lemma}
\smallskip

\textbf{Proof.} We may assume that the subspace $Z_0$ is not
symplectic (since otherwise the assertion of the lemma is trivial).
\smallskip

By Lemma 6 in \cite[Appendix 6.2]{BRC} there exists a symplectic
basis $\{e_i, h_i\}$ in $Z$ such that $\{e_i,
h_i\}_{i=1}^d\cup\{e_i\}_{i=d+1}^p$ is a basis in $Z_0$. To prove the lemma it suffices to show that
the set of vectors $\{Ke_i, Kh_i\}_{i=1}^d\cup\{Ke_i\}_{i=d+1}^p$
can be extended to a symplectic basis in $Z$. By the property of the
map $K$ the set $\{Ke_i, Kh_i\}_{i=1}^d$ is a symplectic basis in
the linear hull $Z_d$ of this set and
$\{Ke_i\}_{i=d+1}^p\subset [Z_d]^{\perp}$. So, we have to find a set
of vectors $\{\tilde{h}_i\}_{i=d+1}^p$ in $[Z_d]^{\perp}$ such
that $\Delta(Ke_i,\tilde{h}_j)=\delta_{ij}$, $i,j=\overline{d+1,p}$.
This set can be constructed sequentially: the vector
$\tilde{h}_{d+1}$ can be chosen in the subspace
$$[Z_d]^{\perp}\cap\left[\left[\{Ke_i\}_{i=d+2}^p\right]^{\perp}\backslash\left[\{Ke_i\}_{i=d+1}^p\right]^{\perp}\right],$$
the vector $\tilde{h}_{d+2}$ -- in the subspace
$$[Z_{d+1}]^{\perp}\cap\left[\left[\{Ke_i\}_{i=d+3}^p\right]^{\perp}\backslash\left[\{Ke_i\}_{i=d+2}^p\right]^{\perp}\right],$$
where $Z_{d+1}=Z_d\oplus\mathrm{lin}[Ke_{d+1},\tilde{h}_{d+1}]$, etc.
$\square$

\bigskip

I am grateful to A.S.Holevo for useful discussion.

\end{document}